# Physics-driven innovations toward the democratization of proton therapy

*Vivek Maradia[1], Martin Bues[2]*

[1]Department of Radiation Oncology, Stanford University School of Medicine, Stanford, CA, USA

[2]Department of Radiation Oncology, Mayo Clinic, Phoenix, AZ 85054, USA

**Abstract**

Proton therapy exploits the finite range of charged particles in tissue to achieve dose distributions no photon-based modality can replicate. Yet the modality reaches fewer than 1% of patients who might benefit—a gap rooted in cost and complexity rather than clinical evidence. This Review reframes proton therapy adoption as a physics problem. Two fundamental bottlenecks are identified: cost, arising from scaling laws governing accelerator design, beam transport, and radiation shielding; and motion, arising from the spatiotemporal mismatch between sequential pencil beam scanning and respiratory tumour displacement. We trace how successive compact architectures—from gantry-integrated energy selection to gantry-mounted accelerators and upright fixed-beam systems—have progressively reduced facility scale toward LINAC-like simplicity and cost-effectiveness. An economic physics framework incorporating fixed and variable operating costs demonstrates that delivery speed has greater leverage on cost per patient than capital cost reduction alone. Field delivery times of approximately 10 seconds—now demonstrated across fundamentally different architectures—simultaneously suppress the interplay effect and enable the patient throughput required for financial viability. The same physics that resolves the motion problem drives the economic case for broad adoption. Emerging directions including proton arc therapy, FLASH irradiation, and adaptive delivery define the path toward global democratization of the modality.

## 1. Introduction

Proton therapy occupies a unique position in the landscape of cancer radiotherapy, grounded in a physical principle that no photon-based modality can replicate: the finite, controllable range of charged particles in tissue[1]. As protons traverse matter, they deposit the bulk of their energy in a sharply defined distal peak—the Bragg peak—before coming to an abrupt stop[2]. By superimposing multiple such peaks through intensity-modulated proton therapy (IMPT), clinicians can sculpt dose distributions of extraordinary conformity around tumour volumes while dramatically sparing surrounding healthy tissue[3]. This dosimetric precision has positioned proton therapy as the most spatially selective form of external beam radiotherapy currently available.

Yet translating physical elegance into unequivocal clinical advantage has proved more complex than the physics alone would suggest. Computational and planning studies have consistently demonstrated the superiority of IMPT over photon-based radiotherapy in terms of integral dose to healthy tissue (as shown in Figure 1)[3,4], and emerging evidence points to tangible clinical gains—reduced acute toxicities in head and neck cancer[5], improved outcomes in pediatric populations[6], and lower rates of treatment-induced complications including secondary malignancies[7,8]. Nevertheless, the magnitude of benefit for many adult indications remains an active area of investigation, in part because the most significant advantages of proton therapy are often long-term and second order[7,8]: improvements in quality of life, enhanced tolerance to concurrent chemotherapy[9–11], and avoidance of late effects whose incidence can take years or decades to manifest.

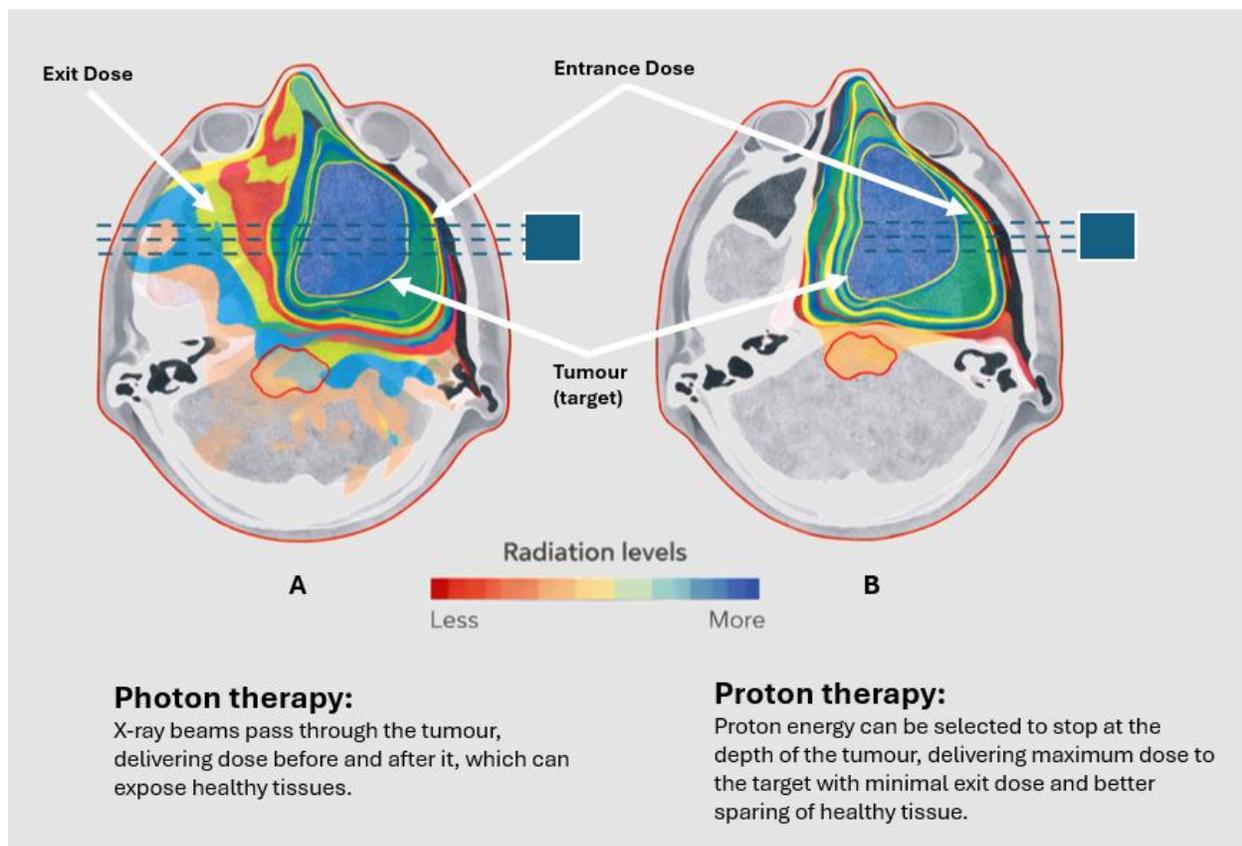

*Figure 1 Illustration of photon and proton radiation beams directed toward a tumour: the photon beam passes through the target, delivering entrance and exit dose, while the proton beam deposits maximum energy at the tumour depth and stops, minimizing exposure to surrounding healthy tissue.*

This nuanced clinical picture sits at the centre of an ongoing debate. Randomised controlled trial evidence directly comparing proton therapy with photon radiotherapy is limited; completed and ongoing trials—including comparisons in breast, lung, and head and neck cancers—have in many cases failed to demonstrate statistically significant differences in primary endpoints within the timeframes studied[12–15]. A notable exception is a recent study in *The Lancet* that demonstrated a significant survival benefit for patients with oropharyngeal cancers who were treated with proton therapy (IMPT)[16] — yet even this result has not resolved the cost-effectiveness debate, because the clinical case for proton therapy cannot be separated from its economics.

Critics argue that even such benefits are too long-term and too small in absolute terms to justify the cost premium[17]. This Review takes a different view: the clinical comparison is inseparable from the cost comparison, and the cost comparison is now changing. A treatment with a superior dose distribution but a prohibitive cost will always lose a cost-effectiveness comparison—not because the physics is wrong, but because the economics are wrong. The central argument of this Review is that the economics are now changing faster than the clinical debate has recognized, and that as they do, the calculus of which patients benefit from proton therapy expands with them.

The physical requirements of particle acceleration, beam transport, and dose delivery impose capital and operational expenditures that dwarf those of conventional radiotherapy[17,18]. This economic burden is not independent of the clinical debate—it is central to it. The case for broad proton therapy adoption transforms entirely if it can be delivered at substantially lower expense, and there are compelling reasons to believe that such reductions are now within reach.

The history of proton therapy is best understood as a sequence of conceptual phase transitions rather than incremental refinements. Its intellectual foundations were laid by Robert R. Wilson in 1946, who recognised that the Bragg peak could be exploited therapeutically and outlined the essential elements of a delivery system[2]. Early clinical implementations in the 1950s at Lawrence Berkeley and subsequently Harvard demonstrated feasibility using physics research accelerators, but were limited by the constraints of passive scattering delivery—broad fields shaped by hardware, with unnecessary dose deposited proximal to the target[19,20]. The introduction of CT imaging enabled model-based treatment planning and accurate Bragg peak placement[21]. The opening of the first hospital-based centre at Loma Linda in 1990 established the multi-room gantry architecture that would define proton therapy for two decades[22,23]. The most consequential delivery innovation was pencil beam scanning (PBS), pioneered by Pedroni at PSI: replacing passive hardware with active magnetic deflection of a narrow beam, enabling IMPT and fundamentally transforming dose-shaping capability[3,24]. Today, PBS is the dominant clinical technique worldwide—and the source of the two challenges that define the current era.

From a physics perspective, those challenges reduce to two fundamental and interdependent bottlenecks. The first is cost, which reflects the scaling laws of accelerator physics, beam optics, and radiation shielding, and ultimately determines which patients and health systems can access the modality[25–29]. The second is motion, which arises from the intrinsically sequential nature of PBS delivery and its sensitivity to anatomical movement—an interplay effect that can compromise the very dose conformity that makes proton therapy valuable[30–39]. Together, these constraints define the outer limits of current systems and shape the frontier of innovation (Figure 2).

This Review examines proton therapy through the lens of these twin bottlenecks. §2 establishes their physical basis—the scaling laws that drive cost, and the spatiotemporal dynamics that drive the degradation of the dose distribution caused by motion. §3 introduces an economic physics framework that reveals why delivery speed is the primary lever on cost per patient, motivating the engineering choices that follow. §4 traces the architectural evolution from multi-room facilities to compact single-room systems and upright fixed-beam platforms. §5 analyses the delivery physics strategies—beam current optimization, spot reduction, and energy modulation—that achieve the 10-second delivery times now demonstrated in clinical practice. §6 surveys emerging directions. The central argument throughout is that cost and motion are not separate engineering problems but aspects of a single physical constraint—delivery speed—whose resolution simultaneously unlocks clinical, operational, and economic improvement toward the democratization of the modality.

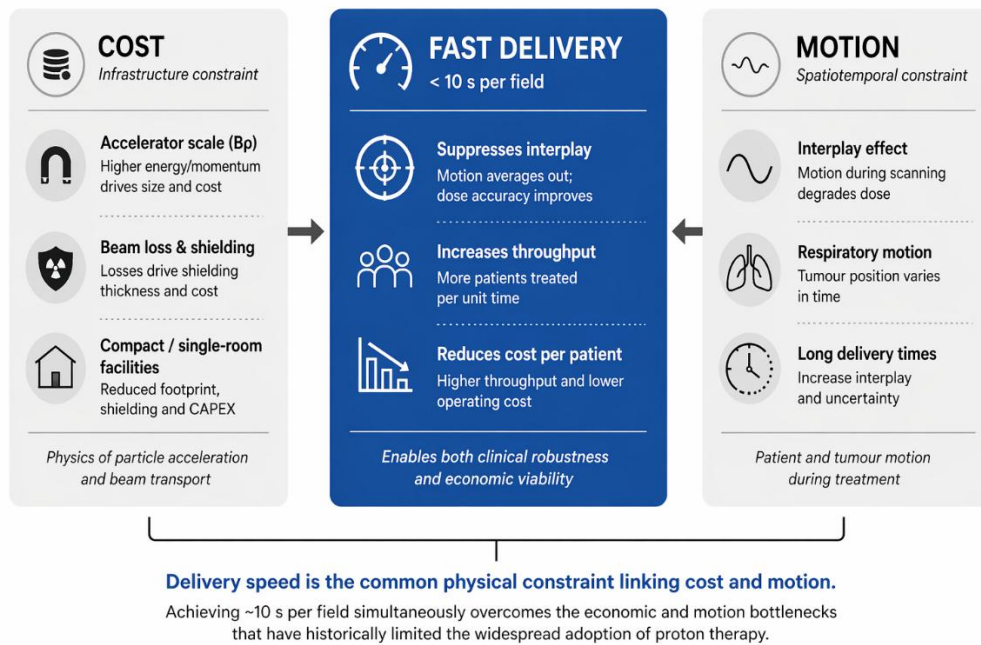

*Figure 2* **The two fundamental bottlenecks in proton therapy and the unifying role of fast delivery.** *Cost, arising from accelerator scaling, beam loss and shielding, and facility infrastructure, and motion, arising from the spatiotemporal mismatch between sequential pencil beam scanning and tumour dynamics, define the principal constraints on current systems. Fast delivery (<10 s per field) occupies the intersection of these domains, simultaneously suppressing the interplay effect and enabling compact single-room implementations with increased throughput and reduced cost per patient.*

## 2. Fundamental limitations: cost and motion

Despite its favorable depth–dose characteristics, proton therapy remains bounded by two fundamental and tightly coupled constraints: cost and motion. These are not merely practical difficulties but reflect deeper physical laws governing how proton beams are generated, transported, and delivered in space and time. Understanding them as physics problems—rather than engineering challenges to be patched incrementally—is essential to identifying viable pathways for meaningful technological advance.

## 2.1 Cost as a scaling law problem: accelerator, beamline, and shielding constraints

The high cost of proton therapy is rooted in the physical requirements of generating and controlling high-energy charged particle beams. Clinically relevant treatment depths require proton kinetic energies spanning 70–230 MeV. To reach the deepest targets, beams must be accelerated to approximately 230 MeV, and at these peak energies the beam's magnetic rigidity—the product of momentum and charge—determines the scale of every downstream component. The fundamental relation $B\rho = p/q$ where $B$ is the magnetic field strength, $\rho$ is the bending radius of the particle's path, $p$ is the particle's momentum and $q$ is the charge, dictates that bending a high-momentum proton beam requires either stronger magnetic fields or larger bending radius (Figure 3(a))[40]. In practice, this translates into rotating gantries that routinely reach diameters of ~10 m and masses exceeding 100 tonnes, imposing substantial mechanical complexity and driving the need for large treatment rooms and their supporting structures[40].

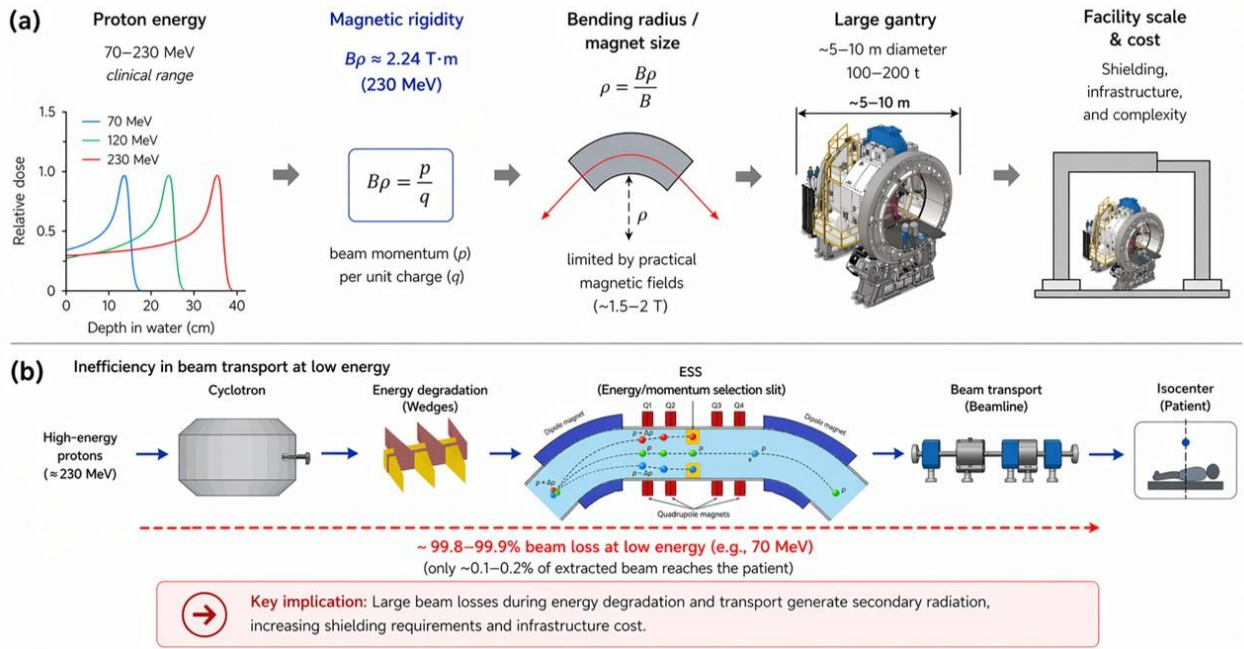

*Figure 3 **Physical origin of system scale and beam transport inefficiency in proton therapy.** (a), Clinical proton energies (70–230 MeV) determine the magnetic rigidity of the beam ($B\rho \approx 2.24$ T·m at 230 MeV), which sets the achievable bending radius for practical magnetic fields (~1.5–2 T). This constraint directly defines gantry size (~5–10 m diameter, 100–200 t), facility scale, and overall system cost[18]. (b), Schematic of cyclotron-based beam delivery illustrating the physical origin of beam transport inefficiency. A fixed high-energy beam (~230 MeV) is degraded to clinical energies using material wedges, increasing energy spread and emittance. The downstream energy selection system (ESS) removes off-momentum particles, followed by transport through the beamline to the patient. At low energies (~70 MeV), cumulative losses during degradation and selection reach ~99.8–99.9%, such that only ~0.1–0.2% of the extracted beam reaches the isocenter[41]. These losses generate secondary radiation and drive shielding requirements, establishing a direct link between beam transport inefficiency and infrastructure cost.*

In multi-room facilities, a central accelerator—most commonly a cyclotron/synchrocyclotron, distributes beam to multiple treatment rooms through an extended network of bending magnets, quadrupoles, switching elements, and energy selection components. The spatial extent of this network is governed by the same magnetic rigidity constraints as the gantries, and the resulting facility footprints can approach the scale of a football field. Shielding requirements, driven by secondary neutron production, add further mass and volume[42–44]. The architecture is therefore inherently large, expensive to construct, and costly to operate.

A defining inefficiency of cyclotron-based systems is energy degradation[45–48]. Because a cyclotron extracts beam at a fixed energy, a variable-thickness degrader must reduce this to the clinical treatment energy. Degradation increases energy spread, angular divergence, and transverse emittance, requiring a downstream energy selection system to collimate the beam and restore an acceptable phase space[48]. This process is intrinsically lossy, particularly at low energies where thicker degraders are required: combined losses in the degrader and energy selection system can reach approximately 99.9% of the extracted beam for low-energy beams (70–100 MeV), meaning that for every thousand protons leaving the cyclotron, fewer than one reaches the patient at the correct energy. The particles lost in this process generate an intense field of secondary radiation in the degrader region—predominantly neutrons—that demands thick concrete shielding around the degrader vault. The shielding required to contain this radiation becomes a dominant cost driver in the facility, establishing a direct and quantitative coupling between beam transport inefficiency and infrastructure expenditure[42,44].

The quantitative consequences of these scaling laws are substantial. The magnetic rigidity of a 230 MeV proton beam is $B\rho \approx 2.24$ T·m, requiring dipole fields of 1.5–2 T over bending radii of 1.2–1.5 m in clinical gantries—directly setting the minimum gantry diameter at ~5 m and the structural steel mass at 100–200 tonnes[49]. Shielding thickness scales approximately logarithmically with neutron fluence: reducing beam loss in the degrader region by a factor of ten allows concrete wall thickness to decrease from ~3.5 m to ~2.5 m, reducing shielding mass by hundreds of tonnes per vault (Figure 3)[42,44]. The cost of proton therapy therefore emerges as a multi-scale physical problem spanning accelerator design, beamline phase-space dynamics, radiation shielding, and temporal delivery efficiency. Meaningful cost reduction demands not incremental improvements within this architecture, but a rethinking of the beam delivery paradigm itself—particularly approaches that minimize or eliminate degradation-induced losses and their associated shielding burden.

## 2.2 Motion as a spatiotemporal physics problem

Where cost is primarily a spatial and infrastructural constraint, motion represents a fundamentally spatiotemporal limitation on dose delivery—one that has been acutely exposed by the transition from passive scattering to PBS. Because PBS constructs a three-dimensional dose distribution sequentially over time, it is intrinsically sensitive to any anatomical displacement occurring between the delivery of successive spots or energy layers[39]. In thoracic and abdominal tumours, where respiratory motion continuously modulates both target position and water-equivalent path length, this sensitivity is clinically decisive. A tumour that moves 15–20 mm with each breathing cycle presents a qualitatively different challenge to a sequentially scanned beam than to the broad fields of passive scattering—the very specificity that makes PBS superior for dose shaping also makes it vulnerable to the kinematics of the geometric target it is trying to conform to.

The governing physical phenomenon is the interplay effect: the interaction between tumour motion and the temporal sequence of spot-by-spot, layer-by-layer dose delivery[30–32,34,50–52]. Because individual spots are deposited at different moments across a delivery that may span tens of seconds, the dose actually received by any given tissue voxel depends not only on the planned spot weights but on where that voxel happens to be each time the beam passes through its depth range. The resulting dose distribution is a probabilistic function of the relative phase between beam delivery and breathing motion—it cannot be predicted from anatomy alone and cannot be corrected by a geometric margin. It is a dynamical mismatch between two independent timescales: the ~3–5 second period of the respiratory cycle and the total field delivery time, which in current clinical practice routinely extends from one to several minutes for a single PBS field, and to 30–40 minutes per fraction when respiratory gating is employed for moving targets[53,54].

Current clinical strategies address the interplay effect indirectly by modifying delivery conditions rather than eliminating the temporal mismatch itself. Rescanning delivers dose in multiple sub-fractions so that motion-induced inhomogeneities average across breathing cycles, but each additional pass multiplies delivery time by a factor of four to eight[32,50,55,56]. Respiratory gating restricts delivery to a 30–40% window of the breathing cycle, extending treatment time by a factor of two to three while introducing dependence on surrogate fidelity[37,57]. Breath-hold suspends respiratory motion but is limited by patient breath-hold duration—typically 5–15 seconds—which is often shorter than the time required to deliver a full PBS field, necessitating multiple interrupted deliveries[58–60]. In practice these strategies are frequently combined, extending treatment sessions for thoracic and abdominal tumours to 30–45 minutes per fraction[53,54]. Every current mitigation approach accepts the fundamental constraint of slow delivery and works around it; none eliminates the root cause. The solution is direct: if field delivery could be completed within a single comfortable breath-hold of 5–10 seconds, the interplay effect would be suppressed at source, rescanning would be unnecessary, and gating replaced by a brief breath-hold compatible with nearly all patients. As demonstrated in §5, this is no longer a theoretical aspiration.

Importantly, fast delivery does not introduce new physical error sources or require a fundamentally different QA framework. However, by compressing dose delivery into a short time window, it alters how existing uncertainties propagate and manifest clinically. In particular, motion-induced effects are suppressed, while the relative importance of delivery precision, system synchronization, and dosimetric control increases. This represents a redistribution—not an expansion—of the uncertainty budget.

## 3. Economic physics: why delivery speed is the primary lever

Before examining how the field has responded to the physical constraints described in §2, it is worth establishing what those constraints cost—and why the answer is not obvious. Conventional debate frames proton therapy economics around capital expenditure: a large cyclotron/synchrotron-based facilities are expensive, therefore compact systems are cheaper, therefore the problem reduces to making the accelerator and beamline smaller. This framing is wrong, or at least radically incomplete. As the analysis below shows, the accelerator is a minor cost contributor, the dominant cost levers are governed by beam physics in ways that are counterintuitive, and the single quantity with the greatest leverage on cost per patient is one that conventional cost discussions rarely mention: delivery speed. Establishing this clearly is essential to understanding why the architectural and delivery innovations in §4 and §5 are not merely engineering improvements but economically motivated physical choices.

## 3.1 Where the money goes: a facility cost anatomy

For a representative compact single-room gantry based facility, based on 2019 data[61], with a total project budget of approximately 40 M€, the proton system itself accounts for 50% (~20 M€); the remaining 50% is absorbed by building construction (15%), project financing (15%), land (7%), imaging suite (8%), and software (5%). Within the 20 M€ system price, the accelerator represents only 10%—approximately 2 M€—with transport and installation (27%), beam line and gantry (13%), and manufacturer gross margin (35%) accounting for the bulk. The accelerator therefore contributes just 5% of total facility cost. A benchmark that has emerged from market analysis captures the consequence: a proton accelerator that costs more than 2 M€ to build, assemble, tune, and factory-test was not competitive at the time of this analysis. The dominant cost levers are not accelerator physics but building size—which scales directly with beam losses and shielding requirements—system integration complexity, and manufacturing scale. Crucially, improving beam transport efficiency reduces shielding volume and building cost simultaneously: a tenfold improvement in low-energy transmission, achievable through the optics and material strategies described in §4 and §5, can reduce building cost by 20–40% by permitting thinner shielding vaults. Although absolute figures have risen with construction cost inflation since this analysis, the proportional cost anatomy—and in particular the dominance of building and system integration costs over accelerator cost—remains structurally valid.

## 3.2 Throughput as the primary economic lever

To understand why delivery speed dominates the economics of proton therapy, it is necessary to examine the cost structure carefully. The analysis concerns the structure of CPP rather than its absolute value; the conclusions are therefore robust to the cost inflation observed since the 2019 reference period. The annual operating cost of a proton therapy facility ($C\_op$) is not a single fixed number—it is the sum of two distinct components. The first is the fixed cost ($C\_fixed$): accelerator maintenance, facility overhead, and the minimum clinical staffing required to keep the department running regardless of how many patients are treated. The second is the annual variable cost ($C\_var \cdot N\_day \cdot D\_op$): the incremental staff time per patient per day, accumulated across all $D\_op$ operating days in the year. This decomposition, though straightforward, has a consequence that is not obvious until the full cost per patient (CPP) expression is written down, as formalized in Box 1.

Box 1 reveals three levers on CPP. Reducing $C\_fac$ (capital cost) improves the first term in the numerator but leaves $C\_fixed$ and $C\_var$ unchanged—so halving capital cost reduces CPP by less than 25% for a typical facility where operating cost accounts for more than half of lifetime expenditure. Increasing $N\_day$ (patients per day) reduces

CPP by spreading fixed costs across more patients—doubling N_day roughly halves CPP. And reducing F_mean (fractions per patient) through hypofractionation reduces CPP by increasing the number of distinct patients treated per year on the same machine. System uptime is a further implicit lever: D_op in the CPP expression reflects actual treatment days, so low uptime directly raises CPP through the fixed-cost term. In practice this is not a binding constraint for modern compact proton systems, which have demonstrated LINAC-comparable uptime profiles exceeding 90% over decade-long clinical records[62,63], and D_op is therefore treated as a facility-planning constant in the analysis that follows.

---

**Box 1 | Economic physics of cost per patient** [64,65]

Annual operating cost decomposes as:

$$C_{op} = C_{fixed} + C_{var} \cdot N_{day} \cdot D_{op}$$

Cost per patient (CPP) as:

$$CPP = \frac{C_{fac} \cdot T_{life}^{-1} + C_{op}}{N_{day} \cdot D_{op} \cdot F_{mean}^{-1}}$$

*Variable definitions*

| Symbol | Definition | Units |
|---|---|---|
| C_fac | Total facility capital cost | M€ |
| T_life | System lifetime | years |
| C_fixed | Annual fixed operating cost (staffing, maintenance, overhead) | M€/yr |
| C_var | Variable cost per patient per day (setup, imaging, delivery, QA) | €/pt |
| N_day | Patients treated per day | pt/day |
| D_op | Annual operating days | days/yr |
| F_mean | Mean fractions per patient | fractions |

Key structural property: F_mean and N_day appear only in the denominator. When fast delivery simultaneously triples N_day and reduces C_var by a factor of approximately three, the product C_var·N_day·D_op (i.e. C_op) in the numerator remains approximately constant and CPP reduction is driven entirely by the growing denominator—a compound benefit that capital cost reduction alone cannot replicate.

---

The critical—and non-obvious—insight is what delivery speed does to this equation. Faster delivery does not only increase N_day. It also reduces C_var per patient, because each patient occupies the treatment room for less time. Assuming a fixed patient setup time of 15 minutes per fraction and a 10-hour clinical day, a clinical team that can manage 40 patients per day with 40–50 second total delivery[66,67] (4–5 fields of ~10 seconds each) could only manage approximately 13 with 30-minute delivery on identical staffing (total room time: ~16 vs 45 minutes per patient)[68]. This linear relationship holds across the moderate utilization range typical of a single-room facility; at very high volumes a step increase in staffing introduces a discontinuity, after which the relationship resets at the new staffing level. Within the normal operating range, however, ultra-fast delivery acts simultaneously on both the numerator (reducing C_var) and the denominator (increasing N_day) of the CPP expression—a compounding benefit that capital cost reduction alone cannot produce. For a 40 M€ single-room facility, fast delivery that halves treatment time approximately doubles N_day while halving C_var, reducing CPP by approximately 50% relative to a slow-delivery baseline with identical hardware. Hypofractionation—reducing F_mean from 25 to 10 fractions—contributes a further approximately 60% CPP reduction, and is itself enabled by fast delivery, since higher per-fraction doses require either higher beam current or longer delivery times[17,69]. The two strategies are therefore not independent: fast delivery is the physical prerequisite for both.

## 3.3 Proton versus photon: a narrowing gap

LINAC-based radiotherapy carries facility costs of 5–10 M€ per room, achieving CPP of 3,000–25,000 € depending on technique and utilization—a factor of 3–20 below current compact proton systems. This gap is not static. As compact proton systems approach facility costs of 20–25 M€[61,64] and throughput of 40–50 patients per day, CPP converges toward 20,000–35,000 €. When comprehensive health economic analyses account for downstream cost offsets[70]—reduced acute toxicity management, lower hospitalization rates, and reduced secondary malignancy incidence over decadal horizons[27]—this range falls within cost-effectiveness thresholds[29] for a substantially wider range of indications than proton therapy currently serves. The physics-economics feedback loop accelerates this convergence: ultra-fast beam delivery—achieved through simultaneous optimization of beam-on time, spot transitions, and energy switching, as described in §5—(Table 2) reduces total field delivery time, which increases throughput, which lowers CPP, which expands the eligible patient population, which improves utilization, which further reduces CPP. A system achieving simultaneous 30% reduction in facility cost, 50% improvement in throughput, and 30% reduction in mean fraction number can reduce CPP by approximately 65-80% relative to a conventional baseline—transforming, rather than incrementally improving, the economics of proton therapy.

## 4. Facility-level innovation and architecture

The economic physics framework of §3 establishes that the dominant levers on cost per patient are building size—governed by beam losses and shielding requirements—system integration complexity, and delivery speed. The history of compact proton therapy architecture is the history of progressively relocating the degrader and energy selection system (ESS) closer to the patient, reducing the volume of beamline that must be shielded and the footprint that must be housed. Three distinct architecture classes are now in clinical operation, each representing a different physical strategy for attacking the same problem; the following sections describe each in terms of its governing physics, characteristic performance, and inherent limitations (Table 1). System names are used illustratively to ground the discussion in demonstrated clinical reality, not as comparative evaluations.

Table 1 Compact Proton Therapy Architecture Classes. Each generation advances toward LINAC-like simplicity through a different physical strategy: Gen 1 relocates the energy selection system onto the gantry, reducing facility footprint from ~2500 m² to ~400 m²; Gen 2 eliminates the upstream beamline, raising beam transmission at 70 MeV from ~0.1% to 70–100%; Gen 3 eliminates the gantry itself, approaching LINAC-class capital cost and building volume.

| Parameter | Gen 1: gantry-integrated ESS[71] | Gen 2: gantry-mounted accelerator[72,73] | Gen 3: fixed beam / upright[67,74] |
|---|---|---|---|
| Accelerator type | Superconducting synchrocyclotron | Gantry-mounted synchrocyclotron | Fixed synchrocyclotron or compact synchrotron |
| Accelerator mass | ~50 t | ~15 t | <15 t |
| Facility footprint | ~20 × 20 m | ~12 × 10 m | ~8 × 6 m |
| Energy selection | Gantry-integrated degrader + ESS | Nozzle-based range shifter | Nozzle-based range shifter or direct extraction |
| Beam transmission (70 MeV) | ~0.1% | 70–100% | 70–100% |
| Energy switching time | 0.5–1 s | ~0.5 s | 0.2–2 s |
| Patient position | Supine | Supine | Upright seated / standing |
| Representative example | IBA Proteus ONE | Mevion S250 | Mevion S250-FIT; Hitachi compact syn; P-Cure system |

## 4.1 First generation: compact single-room gantry systems

Conventional multi-room proton therapy facilities are built around a shared isochronous cyclotron (diameter ~430 cm, mass ~200 tonnes) that distributes beam through an extended network of beamlines, switching magnets, and a centralised degrader and ESS to multiple treatment rooms[40]. The resulting facility footprints approach 50 × 50 m—comparable to a football field—and require purpose-built multi-storey structures incompatible with integration into existing hospital buildings. The same holds for the synchrotron-based system. The capital and construction cost of this architecture, and the minimum patient throughput required to justify it, has historically confined proton therapy to large academic or specialist centers.

The first generation of compact single-room systems addressed this by two simultaneous changes: replacing the large isochronous cyclotron with a superconducting synchrocyclotron[18] (diameter ~250 cm)—roughly a factor of 1.7 reduction in accelerator diameter and a substantial reduction in mass—and, critically, relocating the degrader and ESS from a shared upstream vault to the entrance of the rotating gantry itself[71,75]. In this gantry-integrated ESS architecture, the beam extracted at a fixed energy of 230 MeV travels only a short extraction beamline before reaching the gantry-mounted degrader, where energy selection occurs. Because collimator, and momentum slits rotate with the gantry rather than residing in a fixed shared vault, the radiation environment and shielding burden are contained within a single room. The result is a facility footprint of approximately 20 × 20 m—the scale of a tennis court—small enough to be housed within a conventional hospital building. This single architectural decision—moving the ESS onto the gantry—is what made single-room proton therapy in existing hospitals physically possible.

The beam transport physics of this architecture otherwise resembles that of conventional remote-ESS systems, and its transmission characteristics reflect this. Measured cyclotron-to-isocenter efficiencies are approximately 0.21% at 70 MeV, rising to 12.07% at 230 MeV, with losses concentrated at the extraction quadrupoles (~18% combined), the degrader (~8–34% by energy), and the collimator (~36–45%)[75]. The compactness of the layout means all radiation sources are co-located within the single vault rather than distributed across separate shielded areas, which introduces a specific challenge: concrete activation is more severe in a compact single-room gantry system than in an equivalent multi-room facility, requiring careful shielding design even though the total shielded volume is dramatically smaller[42,43,75].

## 4.2 Second generation: gantry-mounted accelerator systems with direct energy selection

The first-generation compact gantry system demonstrated that a single-room hospital-compatible footprint was achievable, but retained the fundamental physics of the remote degrader: beam losses of ~99.8% at 70 MeV, secondary radiation concentrated in the vault, and energy switching times of 0.5–2 seconds governed by the upstream magnetic beamline. The second generation of compact systems addressed these limitations by eliminating the upstream beamline entirely[72]. The governing physical insight is that if the accelerator is co-located with the treatment position—mounted directly on the rotating gantry—no shared beamline is required, and the degrader can be replaced by a compact nozzle-integrated range shifter operating on mechanical rather than magnetic timescales[72].

In this gantry-mounted synchrocyclotron architecture, a superconducting synchrocyclotron of approximately 15 tonnes and 1.8 m diameter is mounted directly on the rotating gantry. The extracted 230 MeV beam passes immediately into the treatment nozzle, which integrates scanning magnets, a nozzle-based range shifter, real-time dosimetry, and a proton multileaf collimator in a single compact assembly[76]. Because no upstream beamline exists, beam transmission from the accelerator to the patient is 70–100%, and the accelerator can be operated at clinical currents of 1–10 nA—precisely matched to therapeutic need—rather than at the 300–1,000 nA required by isochronous cyclotron systems to compensate for degrader losses[73]. The factor-of-17 reduction in accelerator mass relative to a conventional isochronous cyclotron reflects the progressive application of superconducting magnet technology to achieve the field strengths required for clinical beam energies in progressively smaller volumes.

The nozzle-based range shifter achieves energy layer switching of approximately 0.5 seconds through mechanical plate insertion, rather than the magnet current ramping that limits conventional remote-ESS systems. A further physical consequence of this approach is the generation of invariant Bragg peaks: because the extraction beam is degraded mechanically at a fixed starting energy, Bragg peak width remains nearly constant at 8.2 ± 0.2 mm across the full clinical energy range, reducing the number of energy layers needed for a given spread-out Bragg peak (SOBP)[77]. The principal dosimetric trade-off of in-nozzle energy modulation is an increase in native spot size at low energies due to multiple Coulomb scattering in the range-shifting material—reaching approximately 14 mm penumbra at 70 MeV without collimation[73,77]. Dynamic proton multileaf collimation, providing automated layer-by-layer aperture shaping, reduces this to approximately 3 mm, with effective spot sizes as small as 1.8 mm at 230 MeV. This collimation approach introduces a time overhead per energy layer but removes the need for patient-specific hardware[76].

## 4.3 From gantry-based to gantry-less: transition from supine to upright proton therapy

Both the first- and second-generation architectures retain a rotating gantry for beam angle selection. The gantry is the largest, heaviest, and costliest mechanical component in any proton therapy facility, and its elimination opens a third pathway to compactness. In gantry-less systems, the beam direction is fixed and the patient is repositioned using a robotic couch or upright support system—replacing gantry rotation with patient motion, which is mechanically simpler and cheaper but demands precise reproducible positioning and robust image guidance[78]. Two physical strategies have entered clinical development.

The first retains the gantry-mounted synchrocyclotron of the second generation but removes the rotation mechanism, leaving a fixed horizontal beam modulated by the same nozzle-based range shifter[67,79]. Facility footprint and capital cost approach those of photon LINAC installations. The fixed-beam geometry enables a permanent beam stop for in-vivo range verification and volumetric CT at isocenter for adaptive workflows—capabilities geometrically impossible with a rotating gantry. The patient is treated in an upright seated or standing position, constraining accessible tumour sites to those for which upright positioning is clinically and anatomically acceptable.

The second strategy replaces the cyclotron-based accelerator with a compact synchrotron that extracts beam directly at the required clinical energy without any degrader, eliminating degrader-induced beam loss and secondary radiation entirely[74]. The vault footprint is comparable to a LINAC bunker. The trade-off is a sharply defined, narrow Bragg peak—governed by the beam's intrinsic momentum spread rather than degrader straggling—that requires more energy layers and more spots to cover a given target volume than the broader invariant peaks of nozzle-modulated systems. Synchrotron energy switching is also slower than mechanical plate insertion, both effects increasing delivery time for the same target relative to second-generation architectures. The patient is again treated upright. Across both gantry-less strategies, the long-term dosimetric and clinical implications of upright positioning remain an active area of investigation.

## 4.4 Throughput, reliability, and the LINAC-like operating model

Across all three architecture classes, the ultimate test of a compact proton system is not technical performance in isolation but the ability to sustain high clinical throughput with the reliability and operational simplicity of a linear accelerator. Conventional proton therapy facilities, with their complex multi-component beamlines and shared accelerator infrastructure, have historically required dedicated accelerator physicists and engineers for day-to-day operation—a staffing overhead that adds substantially to effective operating cost and limits deployment to institutions with specialist technical capacity[17]. Compact single-room systems of all three architecture classes have been designed from the outset for operation by clinical radiation therapy staff without dedicated accelerator operators, aligning the operational model with that of a LINAC department[62,63,71]. Decade-long clinical operating records at multiple sites have demonstrated that this is achievable in practice, with availability profiles approaching those of photon LINAC systems[62,63]. This reliability—not merely the physics of fast delivery—is a prerequisite for the patient throughput on which the economic case for compact proton therapy depends[65].

The path to proton therapy at LINAC-like cost per patient does not require LINAC-like capital cost, but it does require LINAC-like delivery speed, LINAC-like reliability, and LINAC-like operational simplicity—all of which flow from physics-level architectural decisions made at the design stage. Compact architecture sets the ceiling on achievable delivery speed; the hardware and algorithmic strategies of §5 show how close to that ceiling current systems can operate in clinical practice.

## 5. Treatment delivery innovation: reducing field delivery time in PBS proton therapy

Improving treatment efficiency is essential to maximizing beam utilization and patient throughput in modern proton therapy facilities and becomes especially acute for moving targets. In lung and liver tumours, where respiratory motion necessitates gating or rescanning, total delivery times routinely extend to 30–40 minutes per fraction[53,54]. Such prolonged treatments not only constrain clinical scalability but compound intra-fraction motion uncertainties, as anatomical changes accumulating over tens of minutes can undermine the very dose conformity that proton therapy is designed to achieve. Reducing total field delivery time is therefore a defining technical objective for the current generation of proton therapy systems—one with simultaneous implications for motion management, patient throughput, and cost. The total delivery time of a proton therapy field divides into two categories: beam-on time, during which dose is actively deposited, and dead time, comprising energy layer switching and spot transition overhead[66]. These three components are physically distinct and require targeted optimization through different approaches. Meaningful reduction in overall treatment duration requires their simultaneous optimization—improving any single component in isolation simply displaces the bottleneck rather than removing it.

*Table 2 Decomposition of PBS field delivery time. Values are for a representative 1-litre target volume. Because the three components arise from independent physical mechanisms, sub-10-second total delivery requires their simultaneous optimisation.*

| Component | Physical origin | Unoptimised (s) | Demonstrated optimised (s) | Primary optimisation approach |
|---|---|---|---|---|
| *Beam-on time* | Beam loses in ESS and along the beamline | 20–60 | 5–8 | High-transmission optics (low-Z degrader, asymmetric collimation, large momentum acceptance)[45,46,80–84] |
| *Total spot transition time* | Scanning magnet repositioning between spots and safety checks | 5–20 | 2–3 | Spot-reduction algorithms; continuous / line scanning; nearest-neighbour sequencing[67,85–88] |
| *Total energy switching time* | Accelerator / ESS energy change between layers | 5–40 | 0.5–2 | Ridge filters; local energy degraders; large momentum acceptance beamlines; multi-energy extraction[89–100] |
| *Total field delivery time* | Sum of all three components | 30–120 | ~10 | Simultaneous optimisation of all three; demonstrated at PSI Gantry 2 and Stanford Mevion S250-FIT[41,66,67] |

## 5.1 Beam-on time: increasing current at the isocenter

Beam-on time is directly proportional to the dose per fraction and inversely proportional to the achievable beam current at the isocenter. Clinical constraints—including minimum monitor unit requirements per spot and the need to maintain beam quality across the full clinical energy range—converge on a target of approximately 5–10 nA at the patient location. Reaching this target in gantry-based remote-ESS systems requires overcoming the substantial transmission losses incurred during energy degradation. Three strategies have been validated to raise isocenter current toward this range.

The first is degrader material optimization. Low-scattering materials such as boron carbide and beryllium generate less multiple Coulomb scattering than conventional graphite, limiting emittance growth at the degrader and improving downstream transport efficiency[46,48]. The second is asymmetric collimation. Conventional circular collimators select equal emittances in both transverse planes, but the alternating focusing of quadrupole and dipole elements in a clinical beamline creates inherently different acceptance limits in the horizontal and vertical planes. Asymmetric collimators that independently optimize beam size and divergence selection in each plane exploit this structural asymmetry, increasing phase-space transmission through the energy selection system by factors of 5–6 for low-energy beams[45]. The third strategy is large momentum acceptance beamlines[95], which relax the precision required of momentum selection and allow a wider phase-space volume to reach the patient, combined with momentum cooling—an emittance-exchange technique in which a wedge-shaped low-Z absorber replaces the conventional ESS slit, reducing the beam momentum spread without discarding particles and enabling order-of-magnitude improvements in transmission for the lowest clinical energies[80]. Applied in combination, these approaches bring beam-on times to approximately 5–8 seconds for target volumes of approximately one liter[41].

## 5.2 Spot transition time: planning and delivery strategies

Spot transition dead time accumulates across the full set of lateral beam positions in an IMPT field. Although individual transitions take only a few milliseconds, their cumulative contribution is significant across the thousands of spots in a clinical plan. Two classes of intervention directly reduce this component.

At the treatment planning level, spot-reduction algorithms iteratively remove low-weight spots while maintaining dosimetric quality constraints, consistently reducing spot counts by 30–50% with equivalent or improved target coverage and organ-at-risk sparing[85]. At the delivery level, line-scanning and continuous scanning approaches replace discrete spot irradiation with a continuously moving beam, eliminating lateral transition events and reducing spot transition time to approximately 2–3 seconds per field in both cyclotron- and synchrotron-based implementations[86]. Continuous scanning is already being implemented in the latest synchrotron-based systems[87,88]. Scanning sequence optimization using nearest-neighbors or travelling-salesman algorithms minimizes total magnet travel distance, providing an additional reduction in repositioning overhead. Synchrocyclotron-based systems offer a structural advantage in this regard: the pulsed beam structure, with beam-off intervals of approximately 1–1.3 ms between extraction pulses, allows scanning magnets to reposition asynchronously during beam-off intervals without contributing to delivery time, effectively decoupling spot transition overhead from the time budget[67]. Through these combined approaches, total spot transition time can be confined to 2–3 seconds per field.

## 5.3 Energy layer switching time: layer reduction and fast modulation

Energy switching time varies from approximately 0.5 seconds in direct-ESS systems to 2 seconds or more in conventional facilities, depending on accelerator type[95,101]. This variability reflects fundamentally different physical constraints: mechanical range shifter plates operate on timescales set by motor dynamics, while magnetic energy selection in remote-ESS systems is governed by magnet inductance and beamline response, and synchrotron energy ramping is limited by ring magnetic field cycling. The most broadly applicable strategy for reducing cumulative energy switching dead time is to reduce the number of energy layers required to cover the target, rather than the per-layer switching speed alone. In multi-energy extraction from a synchrotron, energy switching times can be as low as 0.2 seconds[98–100].

Ridge filters achieve this by broadening the individual Bragg peak depth-dose distribution, allowing a single energy setting to span a greater depth range and directly reducing the number of layer transitions required[90–92,102,103]. Multiple ridge filter designs have been developed and validated across proton and ion beam systems, including mini-ridge filters, two-dimensional ripple filters, and dynamic variants in which broadening can be modulated within a single delivery[89]. Dynamic ridge filters apply maximum broadening where it is dosimetrically safe—interior target voxels—while reverting to minimal broadening for peripheral and distal layers where sharp dose fall-off is required to spare adjacent structures, removing the historical trade-off between delivery speed and dose conformity. Local energy degraders positioned near the patient in the nozzle complement ridge filters by decoupling fine energy modulation from the upstream beamline response, enabling effective switching times below 0.5 s even in facilities where full upstream energy changes require several seconds[72,73]. Large momentum acceptance beamlines further reduce the precision and time required per switching step[95–97].

## 5.4 Combined optimization: clinical translation

Integrating these hardware and algorithmic optimizations across all three delivery time components, total field delivery times of approximately 10 seconds become achievable even for large tumour volumes. Such rapid delivery enables treatment within a single breath-hold, eliminating the need for time-consuming respiratory gating and substantially improving patient throughput. At this timescale, the interplay effect is physically suppressed rather than statistically managed: dose delivery is completed before meaningful anatomical displacement accumulates, transforming breath-hold from a demanding multi-interruption procedure into a straightforward clinical routine accessible to nearly all patients, including those with compromised pulmonary function.

Fast delivery approaches combining different accelerator architectures and planning techniques have been experimentally validated across two representative clinical platforms. At PSI's Gantry 2—a conventional gantry-based cyclotron system—combined high-transmission beam optics and spot-reduced treatment planning demonstrated single-breath-hold delivery for locally advanced NSCLC cases with planning target volumes from 137 to 379 cm³, achieving field delivery times below 10 seconds for conventional fractionation and below 15 seconds for hypofractionation[41,66]. At Stanford Medicine, the Mevion S250-FIT—the most compact upright proton therapy system in clinical operation—demonstrated full-field delivery for lung cancer targets from 100 to 1,000 cc within a 5–10 second breath-hold using software-only optimizations requiring no hardware modifications[67]. Together, these results confirm that sub-10-second delivery is achievable across fundamentally different system architectures, from large gantry-based cyclotron facilities to compact upright proton therapy systems. The broader implication extends directly to economics: reducing treatment time from tens of minutes to seconds increases patients treatable per day, improving machine utilization and lowering cost per fraction—consequences that are analysed quantitatively in §3 (Box 1, Table 2).

## 6. Outlook: converging frontiers

The delivery speeds established in §5 (Table 2) are not an endpoint but an enabler. Each direction surveyed in this section extends fast delivery into a distinct physical dimension: §6.1 examines accelerator concepts that push energy-switching timescales below 10 ms, eliminating the last remaining dead-time bottleneck; §6.2 shows how fast energy modulation unlocks dynamic proton arc therapy, distributing entrance dose over the full gantry hemisphere; §6.3 explores how ultra-high instantaneous dose rates may confer differential biological protection to normal tissue; and §6.4 describes how online adaptation and real-time range verification close the loop between delivery and verification. Together they define a convergent trajectory: more precise, faster, biologically richer, and ultimately more accessible proton therapy.

### 6.1 Next-generation accelerator concepts

Fixed-field alternating gradient (FFA) accelerators use a static magnetic lattice in which beam energy is selected by adjusting injection conditions, rather than by ramping magnetic fields[104,105]. This eliminates the fundamental bottleneck of synchrotron energy switching, enabling energy changes in approximately 10 ms — a factor of ~5 improvement over nozzle-based range shifters and ~50 over remote-ESS systems, sufficient to remove energy switching as a contributor to delivery time entirely. The sub-10-ms switching timescale achievable with FFA architecture is precisely what is required to make dynamic proton arc therapy clinically practical, as the following section describes.

### 6.2 Proton arc therapy

Proton arc therapy distributes entrance dose across a wide angular range by rotating the gantry during delivery, reducing organ-at-risk dose for challenging sites including head and neck and thorax[106,107]. Step-and-shoot discrete arc delivery—gantry pausing at optimized angles—is already in clinical use, with demonstrated reductions in parotid dose of 47–74% and improved normal tissue complication probability (NTCP) versus conventional IMPT[108,109]. Dynamic arc delivery, with the beam on continuously during gantry rotation, remains under development and requires energy switching below 100 ms per layer to be clinically practical. The enabling physical requirement—fast energy modulation at each gantry angle (~0.3 s per layer for a 2-minute arc at 36 angles)—is precisely the capability developed in §5. The continuous delivery geometry of dynamic arc—beam on throughout gantry rotation—also represents the configuration that maximises field-averaged dose rate, making arc delivery a natural platform for FLASH irradiation.

### 6.3 FLASH proton therapy

Delivering dose at rates exceeding ~40 Gy/s may selectively spare normal tissue while maintaining tumour control—the FLASH effect, reproduced in multiple preclinical proton models with transient oxygen depletion the leading mechanistic hypothesis[110]. First-in-human proton FLASH treatment was demonstrated in the FAST-01 trial, with subsequent clinical feasibility studies extending to lung and abdominal targets[111]. A critical unresolved challenge is the absence of a consensus dose-rate definition: instantaneous, spot-averaged, and field-averaged dose rates differ by orders of magnitude for scanned PBS delivery, and the biological threshold of ~40 Gy/s applies to an as-yet undefined quantity—a definitional ambiguity that complicates cross-institutional comparison of preclinical and emerging clinical results[111]. Scanned PBS FLASH requires isocenter currents of 100–200 nA—roughly two orders of magnitude above routine clinical operation—imposing demands on beam optics, scanning magnets, and dosimetry systems that saturate at conventional dose rates. Ridge filter-based delivery raises the effective field-level dose rate by concentrating more dose per energy setting[111].

### 6.4 Online adaptation, range verification, and global access

Online adaptive workflows acquiring daily images at the treatment position and re-optimizing plans before delivery—now achievable in under 60 seconds with AI-driven dose approximation—correct for the anatomical drift that undermines dose conformity over multi-fraction courses[112,113]. Prompt gamma cameras, ion acoustic detectors, and proton CT are advancing toward millimeter-scale in-vivo range verification, with current clinical implementations achieving 2–4 mm precision and research systems approaching 1–2 mm under favorable conditions, transforming range uncertainty from a planning margin into a measurable signal[18,78]. Together these define the path toward closed-loop proton therapy. The access challenge is a separate but equally pressing dimension. Of over 120 clinical proton centres worldwide, virtually none are located in sub-Saharan Africa, few located fewer than five centers in South and Southeast Asia, or Latin America—regions accounting for the majority of global cancer deaths. A single-room system at 20–25 M€ with LINAC-like staffing falls within the capital envelope of tertiary cancer centres in middle-income countries. The scaling laws that make proton therapy expensive are universal; the innovations reducing cost in high-income settings reduce it proportionally everywhere.

## 7. Toward democratization: a physical roadmap

Proton therapy began as a physicist's insight about the Bragg peak. The question it now faces is not whether it works but whether it can reach the patients who need it—approximately 50% of cancer patients require radiotherapy, yet in most low- and middle-income countries even basic radiotherapy access is severely limited. The physics described in this Review defines a credible pathway: a facility that fits in a hospital building, costs 20–25 M€, and operates without a dedicated accelerator staff is not an aspiration—it exists today. The central argument is that delivery speed is the primary lever: it suppresses the interplay effect, raises throughput, enables hypofractionation, and compounds across every patient over a facility's lifetime. The physics is tractable, the engineering is advancing, and the clinical demonstrations are accumulating. What remains is the commitment—from physicists, engineers, clinicians, and health systems—to translate what the physics now enables into what patients everywhere receive.